\newif\ifonecolumn
\onecolumntrue 
\ifonecolumn
  \documentclass[journal,draftcls,onecolumn, 12pt]{IEEEtran}
\else
  \documentclass[journal]{IEEEtran}
\fi
\usepackage[T1]{fontenc}
\usepackage{cite}
\usepackage{array}
\usepackage{stfloats}
\usepackage{url}
\usepackage{amsmath,amssymb}
\usepackage{epsfig,fancyhdr}
\usepackage{CJKutf8}
\usepackage{times}
\usepackage{graphicx}

\newcommand{\bm}[1]{\mbox{\boldmath $#1$}}
\newcommand{\bms}[1]{\mbox{\footnotesize\boldmath$#1$}}
\newcommand{\bmss}[1]{\mbox{\tiny\boldmath$#1$}}

\newcommand{\PAPR}{\mbox{PAPR}}

\newtheorem{thm}{Theorem}

\newtheorem{defin}[thm]{Definition}
\newtheorem{lemma}[thm]{Lemma}
\newtheorem{crly}[thm]{Corollary}
\newtheorem{eg}{Example}
\newtheorem{rmk}{Remark}

\begin{document}
%
\title{Two-Dimensional Golay Complementary Array Sets from Generalized Boolean Functions }
\author{Cheng-Yu~Pai and Chao-Yu~Chen,~\IEEEmembership{Member,~IEEE}
\thanks{The material in this paper was presented in part at the IEEE International Symposium on Information Theory (ISIT), June 2020. This work was supported the Ministry of Science and Technology,
Taiwan, R.O.C., under Grants MOST 109--2628--E--006--008--MY3.}
\thanks{The authors are with the Department
of Engineering Science, National Cheng Kung University, Tainan 701, Taiwan, R.O.C.
(e-mail: \{n98081505, super\}@mail.ncku.edu.tw).}}

\maketitle

\begin{abstract}
  The one-dimensional (1-D) Golay complementary set (GCS) has many well-known properties and has been widely employed in engineering. The concept of 1-D GCS can be extended to the two-dimensional (2-D) Golay complementary array set (GCAS) where the 2-D aperiodic autocorrelations of constituent arrays sum to zero except for the 2-D zero shift. The 2-D GCAS includes the 2-D Golay complementary array pair (GCAP) as a special case when the set size is 2. In this paper, 2-D generalized Boolean functions are introduced and novel constructions of 2-D GCAPs, 2-D GCASs, and 2-D Golay complementary array mates based on generalized Boolean functions are proposed. Explicit expressions of 2-D Boolean functions for 2-D GCAPs and 2-D GCASs are given. Therefore, they are all direct constructions without the aid of other existing 1-D or 2-D sequences. Moreover, for the column sequences and row sequences of the constructed 2-D GCAPs, their peak-to-average power ratio (PAPR) properties are also investigated.
\end{abstract}
\section{Introduction}
One-dimensional (1-D) Golay complementary pair (GCP) \cite{Golay} and its extension, Golay complementary set (GCS) \cite{Tseng72}, have zero autocorrelation sums for non-zero shifts and hence have been found many applications, such as channel estimation \cite{CS_CE}, synchronization \cite{CS_sync}, interference mitigation for multi-carrier code division multiple access (MC-CDMA) \cite{Chen_CDMA}, and peak-to-average power ratio (PAPR) control in orthogonal frequency division multiplexing (OFDM) \cite{Golay_power1,Nee2, Golay_RM, Paterson_00}. Such special 1-D sequence pairs and sets can be extended to two-dimensional (2-D) array pairs and sets, called 2-D Golay complementary array pairs (GCAPs) \cite{Dymond_92,GCAP_Mtsufuji_04, GCAP_Jedwab_07, GCAP_Fiedler_08} and 2-D Golay complementary array sets (GCASs)\cite{GCAP_Zeng_10,Jiang_19}, respectively. For 2-D GCAPs and 2-D GCASs, the aperiodic autocorrelations of constituent arrays sum up to zero except for the 2-D zero shift. Owing to their good autocorrelation properties, they have applications in radar \cite{Weathers_1983}, synchronization \cite{Golomb_1982},\cite{Hershey_1983}, multiple-input multiple-output (MIMO) \cite{Jiang_19}, and can be used as spreading sequences in the 2-D MC-CDMA system \cite{2DCDMA_2004}, \cite{2D_CCC}.

In 1999, Davis and Jedweb first proposed a direct construction of 1-D GCPs based on generalized Boolean functions \cite{Golay_RM}. The construction from generalized Boolean functions has algebraic structure and hence can be friendly for efficient hardware generations. Since then, there have been a number of literature investigating constructions of 1-D sequences from Boolean functions, including GCSs \cite{Paterson_00,AAECC2006,Super_16, Super_172, Super_18,schmidt}, complete complementary codes (CCCs) \cite{MutualGCS_2008,Chen08,Liu_2014,Wu_20}. Z-complementary pairs (ZCPs) \cite{Pai_202,odd-ZCP,even-ZCP,Xie_18,Adhikary_20,Shen_19}, and Z-complementary sets (ZCSs)\cite{Wu_18,Sarkar_19}.

In \cite{Dymond_92}, 2-D binary GCAPs can be obtained from existing 1-D GCPs or 2-D GCAPs by using Kronecker product. Later in \cite{GCAP_Fiedler_08}, Fiedler {\it et al.} applied the method given in \cite{Dymond_92} recursively to construct multi-dimensional GCAPs and then 2-D GCAPs can be obtained via projection. In \cite{GCAP_Mtsufuji_04}, via concatenating existing 1-D GCPs or interleaving existing 2-D GCAPs, 2-D GCAPs can be constructed. In \cite{GCAP_Zeng_10}, Zeng and Zhang proposed a construction of 2-D GCASs based on 2-D perfect arrays and only periodic GCASs were considered. Recently, \cite{Jiang_19} provided constructions of 2-D GCASs from existing 1-D GCSs or 1-D CCCs. In addition to 2-D GCAPs and GCASs, 2-D CCCs were studied in \cite{Luke_85, {2DCDMA_2004}, {2D_CCC},{2D_CCC_CDMA}}. 2-D CCCs can be seen as a collection of 2-D GCASs, where any two different 2-D GCAS are mutually orthogonal. The existing constructions of 2-D CCCs need the help of special 1-D sequences, such as Welti codes \cite{Luke_85} and 1-D CCCs \cite{2D_CCC}.

So far, most constructions of 2-D GCAPs or GCASs still require existing sequences or arrays as kernels. Motivated by this, in this paper, novel constructions of 2-D GCAPs, 2-D GCASs, and 2-D Golay complementary array mate based on generalized Boolean functions are proposed. Our proposed constructions are direct constructions and do not require the aid of any existing arrays or specific 1-D sequences.
The newly proposed 2-D GCAPs and Golay complementary array mates can include our previous results \cite[Th.6]{Pai_20} and \cite[Th.7]{Pai_20} as special cases\footnote{In our previous conference paper \cite{Pai_20}, we provided constructions of 2-D GCAPs and 2-D Golay complementary array mates from Boolean functions which can be found in \cite[Th.6]{Pai_20} and \cite[Th.7]{Pai_20}, respectively. The result from \cite[Th.6]{Pai_20} will be described in Theorem \ref{thm:GCAP} in this paper. Then, we provide more general constructions of 2-D GCAPs and Golay complementary array mates in Theorem \ref{thm:GCAP_general} and Theorem \ref{thm:mate}, respectively.}. Besides 2-D GCAPs and 2-D Golay complementary array mate, we further propose a construction of 2-D GCASs from Boolean functions as well. To the best of authors' knowledge, this is the first work to directly construct GCASs without the aid of other special sequences. Furthermore, we analyze the column sequence PAPR and the row sequence PAPR for our proposed 2-D GCAPs and their PAPR upper bounds are derived, respectively, in this paper. Note that the column sequence PAPR is concerned in the MC-CDMA system \cite{Liu_2014, Li_15}.

The rest of this paper is organized as follows. Section \ref{sec:background} gives some notations and definitions. New constructions of 2-D GCAPs, GCASs, and Golay complementary array mates are presented in Section \ref{sec:GCAS}. The column sequence PAPRs and row sequence PAPRs are also discussed. Finally, we conclude our paper in Section \ref{sec:conclusion}.
\section{Preliminaries and Notations}\label{sec:background}
The following notations will be used throughout this paper:
\begin{itemize}
\item $(\cdot)^*$ denotes the complex conjugation.
\item $(\cdot)^T$ denotes the transpose.
\item $\bm 1$ is an all-one vector.
\item $\mathbb{Z}_q=\{0,1,\cdots,q-1\}$ is the ring of integers modulo $q$.
\item Let $\xi=e^{2\pi \sqrt{-1}/q}$.
\item We consider even integer $q$ in this paper.
\end{itemize}
A complex-valued array ${\bm C}$ of size $L_1\times L_2$ can be expressed as
\begin{equation} \label{eq:complex_array}
     {\bm C} = (C_{g,i}), ~\, 0\leq g< L_1, 0\leq i< L_2.
\end{equation}
\begin{defin}
The 2-D aperiodic cross-correlation function of arrays $\bm C$ and $\bm D$ at shift $(u_1,u_2)$ is defined as
\begin{equation} \label{eq:complex_cross}
\rho({\bm C},{\bm D};u_1,u_2)= \sum\limits_{g=0}^{L_1-1}\sum\limits_{i=0}^{L_2-1}{D}_{g+u_1,i+u_2}{C}^{*}_{g,i},
\end{equation}
where ${D}_{g+u_1,i+u_2}=0$ when $(g+u_1)\not\in \{0,1,\cdots,L_{1}-1\}$ or $(i+u_2)\not\in \{0,1,\cdots,L_{2}-1\}$.
When ${\bm C}={\bm D}$, $\rho({\bm C},{\bm C};u_1,u_2)$ is called the 2-D aperiodic autocorrelation function of ${\bm C}$ and denoted by $\rho({\bm C};u_1,u_2)$. Note that $\rho({\bm C};u_1,-u_2)=\rho^{*}({\bm C};-u_1,u_2)$.
\end{defin}

If we take $L_1=1$, then the array ${\bm C}$ can be reduced to a 1-D sequence ${\bm C}=C_i$ for $i=0,1,\ldots,L_2-1$. Therefore, the corresponding 1-D autocorrelation can be given by
\begin{equation}
  \rho({\bm C};u)=
  \sum\limits_{i=0}^{L_2-1}{C_{i+u}C^{*}_i},
\end{equation}
where $C_{i+u}=0$ when $(i+u)\not\in \{0,1,\cdots,L_{2}-1\}$. In this paper, we only consider $q$-PSK modulation. Thus, we define a $q$-ary array $\bm c$ and (\ref{eq:complex_array}) can be rewritten as
\begin{equation}\label{eq:qpsk_array}
\begin{aligned}
 {\bm c} = (c_{g,i}) \text{ and } {\bm C}=(C_{g,i}) =(\xi^{c_{g,i}})=\xi^{\bm c},
\end{aligned}
\end{equation}
where $0\leq g< L_1$ and $0\leq i< L_2$.
The 2-D aperiodic cross-correlation function given in (\ref{eq:complex_cross}) can also be expressed as
\begin{equation} \label{eq:cross}
\rho({\bm C},{\bm D};u_1,u_2)=\sum\limits_{g=0}^{L_1-1}\sum\limits_{i=0}^{L_2-1}\xi^{d_{g+u_1,i+u_2}-c_{g,i}}.
\end{equation}
If taking $L_1=1$, the corresponding 1-D autocorrelation can be modified as
\begin{equation}
  \rho({\bm C};u)=
  \sum\limits_{i=0}^{L_2-1}\xi^{c_{i+u}-c_i}.
\end{equation}
\begin{defin}[Golay Complementary Set]\cite{Tseng72}\label{defin:1-D_GCS}
 A set of $N$ sequences ${\bm C_{0}},{\bm C_{1}},\cdots,{\bm C_{N-1}}$ of length $L$ is a 1-D GCS, denoted by $(N,L)$-GCS, if and only if
\begin{equation}
 \sum_{l=0}^{N-1}{\rho}({\bm C_{l}};u)=
 \begin{cases}
     0, & u\neq 0 \\
     NL,& u=0.
 \end{cases}
 \label{eq:GCS}
\end{equation}
Note that a GCS is reduced to a GCP by taking $N=2$ and each sequence in a GCP is a Golay sequence.
\end{defin}
\begin{defin}[Golay Complementary Mate]\cite{Tseng72}\label{defin:1-D_mate}
For a GCP $({\bm C_0},{\bm C_1})$, if another GCP $({\bm D_0},{\bm D_1})$ meets the following condition:
\begin{equation}
{\rho}({\bm C_0},{\bm D_0};u)+{\rho}({\bm C_1},{\bm D_1};u)=0~~ \text{for all}~u,
\end{equation}
then they are called the {\em Golay complementary mate} of each other.
\end{defin}
\begin{defin}[Golay Complementary Array Pair]\label{defin:2-D_GCAP}
 A pair of arrays ${\bm C}$ and ${\bm D}$ of size $L_1\times L_2$ is called a GCAP, if
\begin{equation}
 {\rho}({{\bm C}};u_1,u_2) + {\rho}({{\bm D}};u_1,u_2)
  =
 \begin{cases}
     2L_1L_2,& (u_1,u_2)=(0,0)\\
     0, & (u_1,u_2)\neq (0,0).\\
 \end{cases}
 \label{eq:GCAP}
\end{equation}
If $\bm C=(\xi^{c_{g,i}})$ and $\bm D=(\xi^{d_{g,i}})$ where $\bm c=(c_{g,i})$ and $\bm d=(d_{g,i})$ over ${\mathbb Z}_q$ for $0 \leq g<L_1,0 \leq i<L_2$, then we also call this array pair $(\bm c,\bm d)$ a $q$-ary GCAP.
\end{defin}
\begin{defin}[Golay Complementary Array Mate]
Given two GCAPs $({\bm A},{\bm B})$ and $({\bm C,\bm D})$, they are called the {\em Golay complementary array mate} of each other if
\begin{equation}
{\rho}({\bm A},{\bm C};u_1,u_2)+{\rho}({\bm B},{\bm D};u_1,u_2)=0~~ \text{for all}~(u_1,u_2).
\end{equation}
\end{defin}
\begin{defin}[Golay Complementary Array Set]\label{defin:2-D_GCAS}
Given a set of array $G=\{{\bm C}_l|~l=0,\cdots, N-1\}$, where each array ${\bm C}_l$ is of size $L_1\times L_2$, $G$ is called an $(N,L_1,L_2)$-GCAS if
\begin{equation}
\begin{aligned}
&\sum_{l=0}^{N-1}\rho({\bm C}_l;u_1,u_2)=
\begin{cases}
  NL_1L_2,~~~(u_1,u_2)=(0,0)\\
  0,~~~(u_1,u_2)\neq(0,0).
\end{cases}
\end{aligned}
\end{equation}
If ${\bm C}_l=(\xi^{{\bf c}_l})$ where ${\bm c}_l$ is a $q$-ary array over ${\mathbb Z}_q$ for $l=0,1,\cdots, N-1$, then we call the array set $\{{\bm c}_0,{\bm c}_1,\cdots,{\bm c}_{N-1}\}$ a $q$-ary GCAS. Actually, a GCAP is a $(2,L_1,L_2)$-GCAS.
\end{defin}
\subsection{Peak-to-Average Power Ratio}\label{sec:PAPR}
For a $q$-PSK modulated array $\bm C$ given in (\ref{eq:qpsk_array}), we define $ {\bm C}_g$ and $ {\bm C}^T_i$ as the $g$-th row sequence and the $i$-th column sequence, respectively. That is,
\begin{equation}
{\bm C}_g=(C_{g,0},C_{g,1},\cdots,C_{g,L_{2}-1}),~~{\bm C}^T_i=(C_{0,i},C_{1,i},\cdots,C_{L_{1}-1,i})^{T}.
\end{equation}
For the row sequence ${\bm C}_g$, the complex baseband OFDM signal is given by
\begin{equation}
S_{{\bmss C}_g}(t)=\sum_{i=0}^{L_{2}-1}C_{g,i}e^{2\pi\sqrt{-1} i t}=\sum_{i=0}^{L_{2}-1}\xi^{c_{g,i}}e^{2\pi\sqrt{-1} i t},~~0\leq t\leq 1
\end{equation}
where $L_2$ equals to the number of subcarriers. The PAPR of a row sequence ${\bm C}_g$ is defined as
\begin{equation}
\PAPR({\bm C}_g)=\max_{0\leq t\leq 1}\frac{|S_{{\bmss C}_g}(t)|^2}{L_2}
\end{equation}
where $L_2$ is the average power for $q$-PSK modulated sequences. Similarly, the PAPR of a column sequence ${\bm C}^T_i$ can also be given by
\begin{equation}
\PAPR({\bm C}^T_i)=\max_{0\leq t\leq 1}\frac{|S_{{\bmss C}^T_i}(t)|^2}{L_1},
\end{equation}
where
\begin{equation}
S_{{\bmss C}^T_i}(t)=\sum_{g=0}^{L_{1}-1}C_{g,i}e^{2\pi\sqrt{-1} g t}=\sum_{g=0}^{L_{1}-1}\xi^{c_{g,i}}e^{2\pi\sqrt{-1} g t},~~0\leq t\leq 1.
\end{equation}
Note that for an MC-CDMA system, the column PAPR of ${\bm C}^T_i$ is concerned since ${\bm C}^T_i$ is spread in the $i$-th chip-slot over the $L_1$ subcarriers \cite{Liu_2014}.
\subsection{Generalized Boolean Functions}\label{sec:2Dboolean}
Here, we will introduce the 2-D generalized Boolean function. A 2-D generalized Boolean function $f:\mathbb{Z}_2^{n+m}\rightarrow \mathbb{Z}_q$ comprises $n+m$ variables $y_1,y_2,\ldots,y_n,x_1,x_2,\ldots,x_m$, where $x_i,y_g \in \{0,1\}$ for $g=1,2,\ldots,n$ and $i=1,2,\ldots,m$. We define the monomial of degree $r$ as a product of $r$ distinct variables. For example, $x_{1}x_{2}x_{3}$ is a monomial of degree 3 and $x_{1}x_{2}x_{3}y_{1}y_{2}$ is a monomial of degree 5. For simplicity, we define the variables $z_1,z_2,\cdots,z_{n+m}$ as
\begin{equation}\label{eq:z}
z_{l}=
\begin{cases}
y_{l} & \text{for}~1\leq l\leq n;\\
x_{l-n} & \text{for}~n< l\leq m+n,
\end{cases}
\end{equation}
which will be very useful in our proposed construction methods. For a $q$-ary generalized Boolean function with $n+m$ variables, we define the associated array as
\begin{equation}
{\bm f}=
\begin{pmatrix}
f_{0,0}& f_{0,1} & \cdots & f_{0,2^{m}-1}\\
f_{1,0} & f_{1,1} & \cdots & f_{1,2^{m}-1}\\
\vdots & \vdots & \ddots & \vdots   \\
f_{2^{n}-1,0} & f_{2^{n}-1,1} & \cdots & f_{2^{n}-1,2^{m}-1}
\end{pmatrix}\end{equation}
by letting
\begin{equation}\label{eq:boolean}
f_{g,i}=f((g_1,g_2,\cdots,g_n),(i_1,i_2,\cdots,i_m))
\end{equation}
where $(g_1,g_2,\cdots,g_n)$ and $(i_1,i_2,\cdots,i_m)$ are binary representations of the integers $g=\sum_{h=1}^{n}g_{h}2^{h-1}$ and $i=\sum_{j=1}^{m}i_{j}2^{j-1}$, respectively.
\begin{eg}\label{eg:boolean}
If $q=4,n=2$, and $m=3$, the associated array to the generalized Boolean function $f=2z_{1}+z_2+3z_{3}z_{5}+2z_4$ is given by
\[{\bm f}=
\begin{pmatrix}
0& 0 & 2 & 2 & 0& 3 & 2 & 1\\
2& 2 & 0 & 0 & 2& 1 & 0 & 3\\
1& 1 & 3 & 3 & 1& 0 & 3 & 2\\
3& 3 & 1 & 1 & 3& 2 & 1 & 0
\end{pmatrix}.\]
This generalized Boolean function $f$ can also be stated as $f=2y_1+y_2+3x_1x_3+2x_2$ according to (\ref{eq:z}).
\end{eg}
\section{Constructions of GCAPs and GCASs}\label{sec:GCAS}
In this section, constructions of 2-D GCAPs and 2-D GCASs based on generalized Boolean functions will be proposed by each subsection. In addition, a construction of 2-D Golay complementary array mates will be provided as well. Based on our proposed 2-D GCAPs, the column sequence PAPR and row sequence PAPR will be investigated and their upper bounds on PAPRs will also be given, respectively.
\subsection{GCAPs Based on Generalized Boolean Functions}\label{sec:GCAP}
In this subsection, we will first restate a basic construction of 2-D GCAPs in \cite[Th. 6]{Pai_20} and then extend it to new constructions of 2-D GCAPs and 2-D Golay complementary array mates. The proposed constructions can include \cite[Th. 6]{Pai_20} and \cite[Th. 7]{Pai_20} as special cases.

Let us first introduce the well-known constructions of 1-D GCPs \cite{Golay_RM} and 1-D GCSs \cite{Chen08} in the following Lemmas, which will be used hereinafter.
\begin{lemma}\label{lemma:GCP}\cite{Golay_RM,Paterson_00}
For any integer $m\geq 2$, let $\pi$ be a permutation of the set $\{1,2,\cdots,m\}$. Let the generalized Boolean function be given by
\begin{equation}\label{eq:GDJB}
\begin{aligned}
f =  &\frac{q}{2}\sum_{l=1}^{m-1} x_{\pi_{1}(l)}x_{\pi_{1}(l+1)}+\sum_{l=1}^{m}p_{l}x_{l}+p_0
\end{aligned}
\end{equation}
where $p_l \in \mathbb{Z}_q$ for $l=0,1,\cdots,m$. The pair
\begin{equation}\label{eq:GDJ}
(\bm f,\bm f')=\left({\bm f},{\bm f}+\frac{q}{2}{\bm x}_{\pi(1)}\right)
\end{equation}
is a $q$-ary GCP of length $2^m$. Note that such constructed GCP is also called a Golay-Davis-Jedwab (GDJ) pair and its PAPR is at most 2.
\end{lemma}
\begin{lemma}\label{lemma:GCS}\cite{Chen08}
For any integer $m\geq 2$ and any positive integer $k\leq m$, let $\{1,2,\cdots,m\}$ be divided into a partition $I_1,I_2,\cdots,I_k$ and $\pi_{\alpha}$ be a bijection mapping from $\{1,2,\cdots,m_{\alpha}\}$ to $I_{\alpha}$ where $m_{\alpha}=|I_{\alpha}|$ for $\alpha=1,2,\cdots,k$. For the generalized Boolean function given by
\begin{equation}
\begin{aligned}
f= \frac{q}{2}\sum_{\alpha=1}^{k}\sum_{\beta=1}^{m_\alpha-1}x_{\pi_\alpha(\beta)}x_{\pi_\alpha(\beta+1)}+\sum_{l=1}^{m}p_l x_l+p_{0}
\end{aligned}
\end{equation}
where $p_l\text{'s} \in \mathbb{Z}_q$, the set
\begin{equation}\label{eq:GCS_boolean}
\begin{aligned}
&G=\left\{{\bm f}+\frac{q}{2}\sum_{\alpha=1}^{k}\lambda_{\alpha}{\bm x}_{\pi_{\alpha}(1)}:\lambda_{\alpha}\in \{0,1\}\right\}\\
\end{aligned}
\end{equation}
is a $(2^k,2^m)$-GCS. It has been proved that a GCS of size $2^k$ has PAPR at most $2^k$ \cite{Paterson_00}.
\end{lemma}

In what follows, we will provide a theorem to construct 2-D GCAPs based on generalized Boolean functions.
\begin{thm}\label{thm:GCAP}\cite[Th. 6]{Pai_20}
For a $q$-ary array ${\bm c}$, let $\pi_1$ be a permutation of $\{1,2,\cdots,m\}$ and $\pi_2$ be a  permutation of $\{1,2,\cdots,n\}$. Let the generalized Boolean function
\begin{equation}\label{eq:2d}
\begin{aligned}
f =  &\frac{q}{2}\left(\sum_{l=1}^{m-1} x_{\pi_{1}(l)}x_{\pi_{1}(l+1)}+ \sum_{s=1}^{n-1} y_{\pi_{2}(s)}y_{\pi_{2}(s+1)}+x_{\pi_{1}(m)}y_{\pi_{2}(1)}\right)\\
&+\sum_{l=1}^{m}p_{l}x_{l}+\sum_{s=1}^{n}\lambda_{s}y_{s}+p_0
 \end{aligned}
 \end{equation}
 where $p_l,\lambda_{s}\in \mathbb{Z}_q$.
 Then, the array pair
 \begin{equation}
 ({\bm c},{\bm d})=\left({\bm f},{\bm f}+\frac{q}{2}{\bm x}_{\pi_{1}(1)}\right)
\end{equation}
 is a $q$-ary GCAP of size $2^n\times 2^m$.
\end{thm}
\begin{IEEEproof}
Taking a 2-D GCAP $(\bm C,\bm D)$ of size $2^n\times 2^m$, we need to show that
\begin{equation}\label{eq:c+d}
\rho({\bm C};u_1,u_2)+\rho({\bm D};u_1,u_2)=0,~\text{for}~(u_1,u_2)\neq (0,0).
\end{equation}
Let the array
\begin{equation}
{\bm C}=\xi^{\bm c}=
\begin{pmatrix}
\xi^{c_{0,0}}& \xi^{c_{0,1}} & \cdots & \xi^{c_{0,2^{m}-1}}\\
\xi^{c_{1,0}} & \xi^{c_{1,1}} & \cdots & \xi^{c_{1,2^{m}-1}}\\
\vdots & \vdots & \ddots & \vdots   \\
\xi^{c_{2^{n}-1,0}} & \xi^{c_{2^{n}-1,1}} & \cdots & \xi^{c_{2^{n}-1,2^{m}-1}}
\end{pmatrix},
\end{equation}
where $\bm c$ is expressed as
\begin{equation}\label{eq:boolean_c}
\begin{aligned}
{\bm c} =  &\frac{q}{2}\left(\sum_{l=1}^{m-1} {\bm x}_{\pi_{1}(l)}{\bm x}_{\pi_{1}(l+1)}+ \sum_{s=1}^{n-1} {\bm y}_{\pi_{2}(s)}{\bm y}_{\pi_{2}(s+1)}+\frac{q}{2}{\bm x}_{\pi_{1}(m)}{\bm y}_{\pi_{2}(1)}\right)\\
&+\sum_{l=1}^{m}p_{l}{\bm x}_{l}+\sum_{s=1}^{n}\lambda_{s}{\bm y}_{s}+p_0\cdot {\bm 1}.
 \end{aligned}
 \end{equation}
Therefore, (\ref{eq:c+d}) is equivalent to
\begin{equation}
\begin{aligned}
  \rho({\bm C};u_1,u_2)+\rho({\bm D};u_1,u_2)
  &=\sum\limits_{g=0}^{2^{n}-1}\sum\limits_{i=0}^{2^{m}-1}\left(\xi^{c_{g+u_1,{i+u_2}}-c_{g,i}}+\xi^{d_{g+u_1,{i+u_2}}-d_{g,i}}\right)\\
  &=\sum\limits_{i=0}^{2^{m}-1}\sum\limits_{g=0}^{2^{n}-1}\left(\xi^{c_{g+u_1,{i+u_2}}-c_{g,i}}+\xi^{d_{g+u_1,{i+u_2}}-d_{g,i}}\right)\\
  &=0
  \label{eq:c+d_2}
\end{aligned}
\end{equation}
for $-2^{n}<u_{1}<2^{n},-2^{m}<u_{2}<2^{m}$ and $(u_1,u_2)\neq (0,0)$. For given integers $g,i$, let $h=g+u_1,j=i+u_2$. We also let $(g_1,g_2,\cdots,g_n),(h_1,h_2,\cdots,h_n),(i_1,i_2,\cdots,i_m)$, and $(j_1,j_2,\cdots,j_m)$ be the binary representations of $g,h,i$, and $j$, respectively. Then, four cases are considered as follows to prove (\ref{eq:c+d_2}).

{\it Case 1:} We suppose $i_{\pi_{1}(1)}\neq j_{\pi_{1}(1)}$ and $u_2\neq 0$. We can obtain
 \begin{equation}
\begin{aligned}
c_{h,j}-c_{g,i}-d_{h,j}+d_{g,i}=\frac{q}{2}(i_{\pi_{1}(1)}-j_{\pi_{1}(1)}) \equiv \frac{q}{2} \pmod q
\end{aligned}
\end{equation}
implying $\xi^{c_{h,j}-c_{g,i}}/ \xi^{d_{h,j}-d_{g,i}}=-1$ for all $g=0,1,\cdots,2^{n-1}$. Therefore,
\begin{equation}
\sum\limits_{g=0}^{2^{n}-1}\left(\xi^{c_{h,j}-c_{g,i}}+\xi^{d_{h,j}-d_{g,i}}\right)=0.
\end{equation}

{\it Case 2:} Suppose $i_{\pi_{1}(1)}= j_{\pi_{1}(1)}$ and $u_2\neq 0$. We assume $t$ is the smallest number such that $i_{\pi_{1}(t)}\neq j_{\pi_{1}(t)}$. Then, we let $i'$ and $j'$ be integers different from $i$ and $j$, respectively, only in one position, i.e.,  $i^{\prime}_{\pi_{1}(t-1)}=1-i_{\pi_{1}(t-1)},~j^{\prime}_{\pi(t-1)}=1-j_{\pi_{1}(t-1)}$. Hence, we have
 \begin{equation}
\begin{aligned}
{c}_{g,i'} - {c}_{g,i}
& =  \frac{q}{2}\left({i}_{\pi_{1} (t-2)}i^{\prime }_{\pi_{1} (t-1)} - {i}_{\pi_{1} (t-2)}{i}_{\pi_{1} (t-1)}+i^{\prime }_{\pi_{1} (t-1)}{i}_{\pi_{1} (t)}\right.\\
&~~~~~~\left. - {i}_{\pi_{1} (t-1)}{i}_{\pi_{1} (t)} \right) +p_{\pi_{1}(t-1)}i'_{\pi_{1} (t-1)}-p_{\pi_{1}(t-1)}i_{\pi_{1} (t-1)} \\
&\equiv  \frac{q}{2}\left({i}_{\pi_{1} (t-2)}+{i}_{\pi_{1} (t)}\right)+p_{\pi_{1}(t-1)}(1-2i_{\pi_{1} (t-1)}) \pmod {q}.
\end{aligned}
\end{equation}
Due to the fact that $i_{\pi(t-1)}=j_{\pi(t-1)}$ and $i_{\pi(t-2)}=i_{\pi(t-2)}$, we have
\begin{equation}
\begin{aligned}
c_{h,j}-c_{g,i}-c_{h,j'}+c_{g,i'}&\equiv \frac{q}{2}\left({i}_{\pi_{1} (t-2)}-j_{\pi_{1} (t-2)}+{i}_{\pi_{1} (t)}-{j}_{\pi_{1} (t)}\right)\\
&+p_{\pi_{1}(t-1)}(2j_{\pi_{1} (t-1)}-2i_{\pi_{1} (t-1)})\\
&\equiv \frac{q}{2}\left({i}_{\pi_{1} (t)}-{j}_{\pi_{1} (t)} \right)\equiv \frac{q}{2}\pmod {q}
\end{aligned}
\end{equation}
which means $\xi^{c_{h,j}-c_{g,i}}+\xi^{c_{h,j'}-c_{g,i'}}=0.$ Similarly, we can also obtain $\xi^{d_{h,j}-d_{g,i}}+\xi^{d_{h,j'}-d_{g,i'}}=0$ implying
\begin{equation}
\begin{aligned}
&\sum\limits_{g=0}^{2^{n}-1}\left(\xi^{c_{h,j}-c_{g,i}}+\xi^{c_{h,j'}-c_{g,i'}}+\xi^{d_{h,j}-d_{g,i}}+\xi^{d_{h,j'}-d_{g,i'}}\right)=0.
\end{aligned}
\end{equation}

{\it Case 3:} We suppose $h_{\pi_{2}(1)}\neq g_{\pi_{2}(1)}$ and $u_2=0$ implying $j=i$. For simplicity, we let
\begin{equation}
{\bm a}=\frac{q}{2}\sum_{l=1}^{m-1} {\bm x}_{\pi_{1}(l)}{\bm x}_{\pi_{1}(l+1)}~\text{and}~{\bm b}=\frac{q}{2} \sum_{s=1}^{n-1} {\bm y}_{\pi_{2}(s)}{\bm y}_{\pi_{2}(s+1)}.
\end{equation}
Then, (\ref{eq:boolean_c}) can be rewritten as
\begin{equation}\label{eq:2d_m}
{\bm c}={\bm a}+{\bm b}+\frac{q}{2}{\bm x}_{\pi_{1}(m)}{\bm y}_{\pi_{2}(1)}+\sum_{l=1}^{m}p_{l}{\bm x}_{l}+\sum_{s=1}^{n}\lambda_{s}{\bm y}_{s}+p_0\cdot {\bm 1}.
\end{equation}
Thus, we have
\begin{equation}
\begin{aligned}
\sum\limits_{i=0}^{2^{m}-1}\xi^{c_{h,i}-c_{g,i}}&=\sum\limits_{i=0}^{2^{m}-1}\left(\xi^{b_{h}-b_{g}+\frac{q}{2}i_{\pi_{1}(m)}+\sum\limits_{s=1}^{n}\lambda_{s}(h_{s}-g_{s})}\right)\\
&=\left(\xi^{b_{h}-b_{g}+\sum\limits_{s=1}^{n}\lambda_{s}(h_{s}-g_{s})}\right)\sum\limits_{i=0}^{2^{m}-1}\xi^{\frac{q}{2}i_{\pi_{1}(m)}}=0
\end{aligned}
\end{equation}
where the last equality comes from $\sum\limits_{i=0}^{2^{m}-1}\xi^{\frac{q}{2}i_{\pi_{1}(m)}}=0$. Similarly, we can also obtain $\sum\limits_{i=0}^{2^{m}-1}\xi^{d_{h,i}-d_{g,i}}=0$ which means
\begin{equation}
\sum\limits_{i=0}^{2^{m}-1}\left(\xi^{c_{h,i}-c_{g,i}}+\xi^{d_{h,i}-d_{g,i}}\right)=0.
\end{equation}

{\it Case 4:} Suppose $h_{\pi_{2}(1)}= g_{\pi_{2}(1)}$ and $u_2=0$. Assuming $t$ is the smallest integer with $g_{\pi_{2}(t)}\neq h_{\pi_{2}(t)}$, we let $g'$ and $h'$ be the integers different from $g$ and $h$, respectively, only in position $\pi_{2}(t-1)$. That is, $g^{\prime}_{\pi_{2}(t-1)}=1-g_{\pi_{2}(t-1)},~h^{\prime}_{\pi_{2}(t-1)}=1-h_{\pi_{2}(t-1)}$. Using a similar argument as in Case 2, we can obtain
\begin{equation}
\sum\limits_{i=0}^{2^{m}-1}\left(\xi^{{c}_{h,i} - {c}_{g,i}}+\xi^{{c}_{h',i}-{c}_{g',i}}+\xi^{{d}_{h,i} - {d}_{g,i}}+\xi^{{d}_{h',i}-{d}_{g',i}}\right)=0.
\end{equation}

Combining these four cases, we can prove that $(\bm C,\bm D)$ is a 2-D GCAP of size $2^{n}\times 2^{m}$.
\end{IEEEproof}
\begin{crly}\label{crly:row}
The PAPR of any row sequence of the constructed 2-D GCAPs from Theorem \ref{thm:GCAP} is upper bounded by 2.
\end{crly}
\begin{IEEEproof}
We let ${\bm c}_g$ be the $g$-th row sequence of the array $\bm c$ in (\ref{eq:boolean_c}). For a given $g$, ${\bm c}_g$ can be expressed as
\begin{equation}
{\bm c}_g =  \frac{q}{2}\sum_{l=1}^{m-1} {\bm x}_{\pi_{1}(l)}{\bm x}_{\pi_{1}(l+1)}
+\sum_{l=1}^{m}p_{l}{\bm x}_{l}+\frac{q}{2}{\bm x}_{\pi_{1}(m)}g_{\pi_2(1)}+(p_0+\kappa) {\bm 1}
\end{equation}
where
\begin{equation}
\kappa=\frac{q}{2}\sum_{s=1}^{n-1}g_{\pi_{2}(s)}g_{\pi_{2}(s+1)}+\sum_{s=1}^{n}\lambda_{s}g_s
\end{equation}
and $g=\sum_{s=1}^{n}g_s 2^{s-1}$. It can be seen that ${\bm c}_g$ is a sequence of a GDJ pair by Lemma \ref{lemma:GCP}; hence its PAPR is at most 2.
\end{IEEEproof}
\begin{crly}\label{crly:column}
The constructed 2-D GCAPs from Theorem \ref{thm:GCAP} also have column sequence PAPRs at most 2.
\end{crly}
\begin{IEEEproof}
Similar to the proof of Corollary \ref{crly:row}, the $i$-th column sequence ${\bm c}^T_{i}$ is shown as
\begin{equation}
{\bm c}^T_{i}= \frac{q}{2}\sum_{s=1}^{n-1} {\bm y}_{\pi_{2}(s)}{\bm y}_{\pi_{2}(s+1)}
+\sum_{s=1}^{n}\lambda_{s}{\bm y}_{s}+\frac{q}{2}{i}_{\pi_{1}(m)}{\bm y}_{\pi_2(1)}+(p_0+\kappa') {\bm 1}
\end{equation}
where
\begin{equation}
\kappa'=\frac{q}{2}\sum_{l=1}^{m-1}i_{\pi_{1}(l)}i_{\pi_{1}(l+1)}+\sum_{l=1}^{m}p_{l}i_l
\end{equation}
and $i=\sum_{l=1}^{m}i_l 2^{l-1}$. Clearly, each column sequence can be viewed as a sequence of a GDJ pair and therefore has PAPR upper bounded by 2.
\end{IEEEproof}
\begin{eg}\label{eg:row}
For $q=4,~m=3$, and $n=2$, we let $\pi_1=(3,1,2),~\pi_2=(1,2)$, and the generalized Boolean function $f=2(x_3x_1+x_1x_2+y_1y_2+x_2y_1)+x_1$ by taking $p_1=1$. The array pair $(\bm c,\bm d)=(\bm f,\bm f+2{\bm x_{3}})$ given by
\[
{\bm c}=
\begin{pmatrix}
0 & 1 & 0 & 3 & 0 & 3 & 0 & 1 \\
0 & 1 & 2 & 1 & 0 & 3 & 2 & 3 \\
0 & 1 & 0 & 3 & 0 & 3 & 0 & 1 \\
2 & 3 & 0 & 3 & 2 & 1 & 0 & 1 \\
\end{pmatrix}
~\text{and}~
{\bm d}=
\begin{pmatrix}
0 & 1 & 0 & 3 & 2 & 1 & 2 & 3 \\
0 & 1 & 2 & 1 & 2 & 1 & 0 & 1 \\
0 & 1 & 0 & 3 & 2 & 1 & 2 & 3 \\
2 & 3 & 0 & 3 & 0 & 3 & 2 & 3 \\
\end{pmatrix}
\]
is a GCAP of size $4 \times 8$. According to Corollary \ref{crly:row} and Corollary \ref{crly:column}, we know that both the maximum row sequence PAPR  and the maximum column sequence PAPR are bounded by 2. Actually, each row sequence PAPR of $\bm c$ and $\bm d$ is exact 2 and each column sequence PAPR of $\bm c$ and $\bm d$ is 1.7698.

\end{eg}

Next, we extend Theorem \ref{thm:GCAP} to a general construction of 2-D GCAPs.
\begin{thm}\label{thm:GCAP_general}
Let $\pi$ be a permutation of $\{1,2,\cdots,n+m\}$ and the generalized Boolean function can be given by
\begin{equation}\label{eq:2d_general}
\begin{aligned}
f =  &\frac{q}{2}\sum_{l=1}^{n+m-1} z_{\pi(l)}z_{\pi(l+1)}+\sum_{l=1}^{n+m}p_{l}z_{l}+p_0
 \end{aligned}
 \end{equation}
 where $z$ is defined in (\ref{eq:z}) and $p_{l},p_{0}\in \mathbb{Z}_q$. The array pair
 \begin{equation}\label{eq:pair}
 ({\bm c},{\bm d})=\left({\bm f},{\bm f}+\frac{q}{2}{\bm z}_{\pi(1)}\right)
 \end{equation}
 forms a $q$-ary GCAP of size $2^n\times 2^m$.
\end{thm}
\begin{IEEEproof}
Similarly, we need to prove that
\begin{equation}
\begin{aligned}
  \rho({\bm C};u_1,u_2)+\rho({\bm D};u_1,u_2)=\sum\limits_{g=0}^{2^{n}-1}\sum\limits_{i=0}^{2^{m}-1}\left(\xi^{c_{g+u_1,{i+u_2}}-c_{g,i}}+\xi^{d_{g+u_1,{i+u_2}}-d_{g,i}}\right)=0
  \label{eq:auto_sum}
\end{aligned}
\end{equation}
for $-2^{n}<u_{1}<2^{n},-2^{m}<u_{2}<2^{m}$ and $(u_1,u_2)\neq (0,0)$.
From (\ref{eq:qpsk_array}), we know that $\bm C=\xi^{\bm c}$, where $\bm c$ is expressed as
\begin{equation}\label{eq:genreal_boolean_c}
\begin{aligned}
{\bm c} =  &\frac{q}{2}\sum_{l=1}^{m+n-1} {\bm z}_{\pi(l)}{\bm z}_{\pi(l+1)}+\sum_{l=1}^{n+m}p_{l}{\bm z}_{l}+p_0 \cdot {\bm 1}.
 \end{aligned}
 \end{equation}
Then we let $h=g+u_1,j=i+u_2$ for any integers $g$ and $i$. For the sake of easy presentation, we define
\begin{equation}\label{eq:ab}
\begin{aligned}
&a_{l}=
\begin{cases}
g_{l} &\text{for}~ 1\leq l\leq n;\\
i_{l-n} &\text{for}~ n< l\leq m+n,
\end{cases}\\
&b_{l}=
\begin{cases}
h_{l} &\text{for}~ 1\leq l\leq n;\\
j_{l-n} &\text{for}~ n<l\leq m+n.
\end{cases}
\end{aligned}
\end{equation}
Therefore, (\ref{eq:boolean}) can be rewritten as
\begin{equation}
\begin{aligned}
f_{g,i}=f(a_{1},a_{2},\cdots a_{m+n})~~\text{and}~~f_{h,j}=f(b_{1},b_{2},\cdots b_{m+n}),
\end{aligned}
\end{equation}
respectively. Then, we consider two cases to show that (\ref{eq:auto_sum}) holds.

{\it Case 1:} Suppose $a_{\pi(1)}\neq b_{\pi(1)}$. We can obtain
\begin{equation}
c_{h,j}-c_{g,i}-d_{h,j}+d_{g,i}=\frac{q}{2}(a_{\pi(1)}-b_{\pi(1)})\equiv \frac{q}{2} \pmod q
\end{equation}
implying
\begin{equation}
\xi^{c_{h,j}-c_{g,i}}+\xi^{d_{h,j}-d_{g,i}}=0.
\end{equation}

{\it Case 2:} Suppose $a_{\pi(1)}= b_{\pi(1)}$. We assume $t$ is the smallest number such that $a_{\pi(t)}\neq b_{\pi(t)}$. Let $a'$ and $b'$ be integers different from $a$ and $b$, respectively, only in one position. That is, $a'_{\pi(t-1)}=1-a_{\pi(t-1)},~b'_{\pi(t-1)}=1-b_{\pi(t-1)}$. If $1 \leq \pi(t-1)\leq n$, by using (\ref{eq:ab}), we have
\begin{equation}
\begin{aligned}
{c}_{g',i} - {c}_{g,i}
=&\frac{q}{2}\left(a_{\pi(t-2)}g'_{\pi(t-1)}-a_{\pi(t-2)}g_{\pi(t-1)}+ g'_{\pi(t-1)}a_{\pi(t)}\right.\\
&\left. -g_{\pi(t-1)}a_{\pi(t)}\right)+p_{\pi(t-1)}g'_{\pi(t-1)}-p_{\pi(t-1)}g_{\pi(t-1)}\\
\equiv &\frac{q}{2}\left(a_{\pi(t-2)}+a_{\pi(t)}\right)+p_{\pi(t-1)}(1-2g_{\pi(t-1)})\pmod q
\end{aligned}
\end{equation}
where $a'_{\pi(t-1)}=g'_{\pi(t-1)}$ and $a_{\pi(t-1)}=g_{\pi(t-1)}$. Since $a_{\pi(t-2)}= b_{\pi(t-2)}$ and $a_{\pi(t-1)}= b_{\pi(t-1)}$, from (\ref{eq:ab}), we can obtain
\begin{equation}
\begin{aligned}
c_{h,j}-c_{g,i}-c_{h',j}+c_{g',i}&\equiv \frac{q}{2}\left({a}_{\pi (t-2)}-b_{\pi (t-2)}+{a}_{\pi (t)}-{b}_{\pi (t)}\right)\\
&+p_{\pi(t-1)}(2h_{\pi(t-1)}-2g_{\pi(t-1)})\\
&\equiv \frac{q}{2}\left({a}_{\pi (t)}-{b}_{\pi (t)} \right)\equiv \frac{q}{2}\pmod {q}
\end{aligned}
\end{equation}
implying $\xi^{c_{h,j}-c_{g,i}}+\xi^{c_{h',j}-c_{g',i}}=0.$ Similarly, we can also obtain $\xi^{d_{h,j}-d_{g,i}}+\xi^{d_{h',j}-d_{g',i}}=0.$
If $n< \pi(t-1)\leq n+m$, note that $a'_{\pi(t-1)}=i'_{\pi(t-1)-n}$ and $a_{\pi(t-1)}=i_{\pi(t-1)-n}$ according to (\ref{eq:ab}). By following the similar argument as mentioned above, we can get
\begin{equation}
\xi^{{c}_{h,j} - {c}_{g,i}}+\xi^{{c}_{h,j'}-{c}_{g,i'}}+\xi^{{d}_{h,j} - {d}_{g,i}}+\xi^{{d}_{h,j'}-{d}_{g,i'}}=0.
\end{equation}
From Case 1 and Case 2, we can say that the array pair $({\bm C},{\bm D})$ is indeed a 2-D GCAP.
\end{IEEEproof}
\begin{rmk}\label{rmk:GCAP_general}
By setting $\{\pi(1),\pi(2),\cdots,\pi(n)\}=\{1,2,\cdots,n\}$ and $\{\pi(n+1),\pi(n+2),\cdots,\pi(n+m)\}=\{n+1,n+2,\cdots,n+m\}$ in Theorem \ref{thm:GCAP_general}, (\ref{eq:2d_general}) can be rewritten as
\begin{equation}\label{eq:2dr}
\begin{aligned}
f =  &\frac{q}{2}\left(\sum_{l=1}^{n-1} y_{\pi(l)}y_{\pi(l+1)}+ \sum_{l=n+1}^{n+m-1} x_{\pi(l)-n}x_{\pi(l+1)-n+ y_{\pi(n)}x_{\pi(n+1)-n}}\right)\\
&+\sum_{l=1}^{n}p_{l}y_{l}+\sum_{l=n+1}^{n+m}p_{l}x_{l-n}+p_0\\
\end{aligned}
\end{equation}
according to (\ref{eq:z}). It can be observed that (\ref{eq:2dr}) is in the form of (\ref{eq:2d}) and, therefore, Theorem \ref{thm:GCAP_general} includes Theorem \ref{thm:GCAP} as a special case.
\end{rmk}
\begin{rmk}
When comparing (\ref{eq:2d_general}) and (\ref{eq:GDJB}), it can be found that Theorem \ref{thm:GCAP_general} is a generalization of 1-D GDJ pairs to 2-D GCAPs by applying the proper mapping in (\ref{eq:z}). Theorem \ref{thm:GCAP_general} provides an explicit expression of 2-D Boolean functions for 2-D GCAPs.
\end{rmk}

Like Corollaries \ref{crly:row} and \ref{crly:column}, the PAPR properties for the constructed 2-D GCAPs from Theorem \ref{thm:GCAP_general} are described in the following Corollaries.
\begin{crly}\label{crly:row_general}
For a $q$-ary array pair $(\bm c,\bm d)$ from Theorem \ref{thm:GCAP_general}, we define an index set $W=\{l| \pi(l)>n,~l=1,2,\cdots,n+m\}$. If there exists an integer $v$ and nonempty sets $W_{1},W_{2},\cdots, W_{v}$ satisfying the following conditions, then the row sequences of $\bm c~ (\text{or}~\bm d)$ have PAPRs at most $2^v$.
\begin{itemize}
\item[(C1)] $\{W_{1},W_{2},\cdots, W_{v}\}$ is a partition of the set $W$;
\item[(C2)] the elements in each $W_{\alpha}$ are consecutive integers for $1 \leq\alpha\leq v$.
\end{itemize}
\end{crly}
\begin{IEEEproof}
We let ${\bm c}_{g}$ be the $g$-th row sequence of the array $\bm c$. For the ease of presentation, we consider the case for ${\bm c}_{0}$ and the other cases can be obtained by following the similar argument. For simplicity, we let $\sigma_{\alpha}$ be a bijection from $\{1,2,\cdots,m_{\alpha}\}$ to the set $\{\pi(l)-n|l\in W_{\alpha}\}$ with $m_{\alpha}=|W_{\alpha}|$ and $\sigma_{\alpha}(i)=\pi(\min\{W_{\alpha}\}+i-1)-n$ for $\alpha=1,2,\cdots,v$ and $i=1,2,\cdots,m_{\alpha}$. The sequence ${\bm c}_{0}$ can be written as
\begin{equation}\label{eq:row_sequence}
\begin{aligned}
{\bm c}_{0}= \frac{q}{2}\sum_{\alpha=1}^{v}\sum_{\beta=1}^{m_{\alpha}-1} {\bm x}_{\sigma_{\alpha}(\beta)}{\bm x}_{\sigma_{\alpha}(\beta+1)}+\sum_{l=1}^{m}p_{l}{\bm x}_{l}+p_0\cdot {\bm 1},~~~p_{l}\in \mathbb{Z}_q
\end{aligned}
\end{equation}
which lies in a GCS in (\ref{eq:GCS_boolean}). Therefore, from Lemma \ref{lemma:GCS}, we can conclude that the maximum row sequence PAPR of $\bm c$ and $\bm d$ is at most $2^v$.
\end{IEEEproof}
\begin{crly}\label{crly:col_general}
Let $W'=\{l| 1 \leq\pi(l)\leq n,~l=1,2,\cdots,n+m\}$ and follow the similar conditions (C1) and (C2) in Corollary \ref{crly:row_general} with $W$ replaced by $W'$. Then, the array $\bm c~(\text{or}~\bm d)$ from Theorem \ref{thm:GCAP_general} has column sequence PAPR upper bounded by $2^v$.
\end{crly}
\begin{IEEEproof}
Similarly, we let $\sigma_{\alpha}$ be a bijection from $\{1,2,\cdots,n_{\alpha}\}$ to the set $\{\pi(l)|l\in W'_{\alpha}\}$ with $n_{\alpha}=|W'_{\alpha}|$ and $\sigma_{\alpha}(l)=\pi(\min\{W'_{\alpha}\}+l-1)$ for $\alpha=1,2,\cdots,v$ and $l=1,2,\cdots,n_{\alpha}$. The column sequences can be represented as
\begin{equation}\frac{q}{2}\sum_{\alpha=1}^{v}\sum_{\beta=1}^{n_{\alpha}-1} {\bm y}_{\sigma_{\alpha}(\beta)}{\bm y}_{\sigma_{\alpha}(\beta+1)}+\sum_{l=1}^{n}p'_{l}{\bm y}_{l}+p'_0 \cdot {\bm 1},~~~p'_{l}\in \mathbb{Z}_q
\end{equation}
implying that the maximum column sequence PAPR is bounded by $2^{v}$.
\end{IEEEproof}
\begin{crly}\label{crly:number}
The number of distinct $2^n \times 2^m$ array $\bm c$ obtained from Theorem \ref{thm:GCAP_general} is
\begin{equation}
\frac{(n+m)!}{2}\cdot q^{n+m+1}.
\end{equation}
\end{crly}
\begin{IEEEproof}
For a constructed array $\bm c$ from Theorem \ref{thm:GCAP_general}, it can be expressed as
\begin{equation}\label{eq:number}
{\bm c} =  \frac{q}{2}\sum_{l=1}^{n+m-1} {\bm z}_{\pi(l)}{\bm z}_{\pi(l+1)}+\sum_{l=1}^{n+m}p_{l}{\bm z}_{l}+p_0 \cdot {\bm 1}.
\end{equation} We first calculate the number of the quadratic forms
\begin{equation}\label{eq:quadratic}
\frac{q}{2}\sum_{l=1}^{n+m-1} {\bm z}_{\pi(l)}{\bm z}_{\pi(l+1)}.
\end{equation}
Since $\pi$ is a permutation of the set $\{1,2,\cdots,n+m\}$, there exist $\frac{(n+m)!}{2}$ distinct quadratic forms in (\ref{eq:quadratic}). Then, we have $p_{l} \in \mathbb{Z}_q$ for $l=0,1,\cdots,n+m$ and hence we can determine
\begin{equation}
\frac{(n+m)!}{2}\cdot q^{n+m+1}
\end{equation}
different arrays $\bm c$ of size $2^n \times 2^m$ of the form in (\ref{eq:number}).
\end{IEEEproof}
\begin{eg}\label{eg:GCAP_general}
Taking $q=2,~m=3$, and $n=2$, we let $(\pi(1),\pi(2),\pi(3),\pi(4),\pi(5))=(3,4,2,1,5)$. Then, the generalized Boolean function is $f=z_3z_4+z_4z_2+z_2z_1+z_1z_5$ by setting $p_l=0$ for all $l$ in (\ref{eq:2d_general}). Note that $f$ can also be expressed as $f=x_1x_2+x_2y_2+y_2y_1+y_1x_3$ according to (\ref{eq:z}). Then, the array pair $(\bm c,\bm d)=(\bm f,\bm f+{\bm z}_3)$ is a GCAP from Theorem \ref{thm:GCAP_general} where
\[
{\bm c}=
\begin{pmatrix}
0 & 0 & 0 & 1 & 0 & 0 & 0 & 1 \\
0 & 0 & 0 & 1 & 1 & 1 & 1 & 0 \\
0 & 0 & 1 & 0 & 0 & 0 & 1 & 0 \\
1 & 1 & 0 & 1 & 0 & 0 & 1 & 0 \\
\end{pmatrix}
~\text{and}~
{\bm d}=
\begin{pmatrix}
0 & 1 & 0 & 0 & 0 & 1 & 0 & 0 \\
0 & 1 & 0 & 0 & 1 & 0 & 1 & 1 \\
0 & 1 & 1 & 1 & 0 & 1 & 1 & 1 \\
1 & 0 & 0 & 0 & 0 & 1 & 1 & 1 \\
\end{pmatrix}.
\]
For $\bm C=(\xi^{c_{g,i}})$ and $\bm D=(\xi^{d_{g,i}})~\text{with}~0\leq g< L_1, ~0\leq i< L_2$, the aperiodic autocorrelation values of $\bm C$ and $\bm D$ are given in (\ref{eq:R1}) and (\ref{eq:R2}).
\begin{figure*}[ht]
\begin{equation}\label{eq:R1}
\begin{aligned}
&(\rho(\bm C;u_1,u_2))_{u_{1}=-3\sim 3,u_{2}=-7\sim 7}\\
&=\left( {\begin{array}{ccccccccccccccc}
1& 0& 1& 0& 3& 0& -1& 0& -1& 0& -1& 0& -3& 0& 1\\
2& 0& 2& 0& 6& 0& -2& 0& 2& 0& 2&	0& 6& 0& -2\\
3& 0& -1& 0& 1& 0& 1& 0& -3& 0&	1&	0& -1& 0& -1\\
0& 0& 0& 0& 0& 0& 0& 32& 0& 0& 0& 0& 0& 0& 0\\
-1& 0& -1& 0& 1& 0& -3&0& 1& 0&	1&	0& -1& 0& 3\\
-2& 0& 6& 0& 2& 0& 2&0& -2& 0& 6& 0& 2& 0& 2\\
1& 0& -3& 0& -1& 0& -1&0& -1& 0&3 & 0& 1& 0& 1\\
\end{array}}
\right),
\end{aligned}
\end{equation}
\end{figure*}
\begin{figure*}[ht]
\begin{equation}\label{eq:R2}
\begin{aligned}
&(\rho(\bm D;u_1,u_2))_{u_{1}=-3\sim 3,u_{2}=-7\sim 7}\\
&=\left( {\begin{array}{ccccccccccccccc}
-1& 0& -1& 0& -3& 0& 1& 0& 1& 0& 1& 0& 3& 0& -1\\
-2& 0& -2& 0& -6& 0& 2& 0& -2& 0& -2& 0& -6& 0& 2\\
-3& 0& 1& 0& -1& 0& -1& 0& 3& 0& -1& 0& 1& 0& 1\\
0& 0& 0& 0& 0& 0& 0& 32& 0& 0& 0& 0& 0& 0& 0\\
1& 0& 1& 0& -1& 0& 3&0& -1& 0&	-1&	0& 1& 0& -3\\
2& 0& -6& 0& -2& 0& -2&0& 2& 0& -6& 0& -2& 0& -2\\
-1& 0& 3& 0& 1& 0& 1&0& 1& 0& -3 & 0& -1& 0& -1\\
\end{array}}
\right).
\end{aligned}
\end{equation}
\end{figure*}
We can see that their aperiodic autocorrelations sum to zero except for $(u_1,u_2)=(0,0)$.
From Corollary \ref{crly:row_general}, we have $W=\{l|\pi(l)>2\}=\{1,2,5\}$ and let $ W_1=\{1,2\}, W_2=\{5\}$ be a partition of $W$ with $v=2$ implying that the row sequence PAPRs of $\bm c$ and $\bm d$ are at most $2^v=4$. Actually, the maximum row sequence PAPR of $\bm c$ and $\bm d$ is 3.4427. Also, we let $W'=\{l|1 \leq \pi(l)\leq 2\}=\{3,4\}$ which yields $v=1$ and the column sequence PAPR is at most 2 according to Corollary \ref{crly:col_general}. In fact, the column sequence PAPR of each column of $\bm c$ and $\bm d$ is 1.7698.
\end{eg}

Next, a construction of Golay complementary array mates is provided based on Theorem \ref{thm:GCAP_general}.
\begin{thm}\label{thm:mate}
The array pair $({\bm c'},{\bm d'})$ is a Golay complementary array mate of the GCAP $(\bm c,\bm d)$ given in (\ref{eq:pair}) where
\begin{equation}
 ({\bm c'},{\bm d'})=\left({\bm f}+\frac{q}{2}{\bm z}_{\pi(n+m)},{\bm f}+\frac{q}{2}{\bm z}_{\pi(1)}+\frac{q}{2}{\bm z}_{\pi(n+m)}\right)
 \end{equation}
 and $\bm f$ is the associated array to the Boolean function $f$ in (\ref{eq:2d_general}).
\end{thm}
\begin{IEEEproof}
For the 2-D GCAPs $(\bm C,\bm D)=(\xi^{\bm c},\xi^{\bm d})$ and $({\bm C'},{\bm D'})=(\xi^{\bm c'},\xi^{\bm d'})$ of size $2^{n} \times 2^{m}$, we would like to show that
\begin{equation}
\begin{aligned}
  &\rho({\bm C},{\bm C'};u_1,u_2)+\rho({\bm D},{\bm D'};u_1,u_2)=\sum\limits_{g=0}^{2^{n}-1}\sum\limits_{i=0}^{2^{m}-1}\left(\xi^{c'_{g+u_1,{i+u_2}}-c_{g,i}}+\xi^{d'_{g+u_1,{i+u_2}}-d_{g,i}}\right)=0
\end{aligned}
\end{equation}
for $-2^{n}<u_{1}<2^{n}$ and $-2^{m}<u_{2}<2^{m}$.
It is noted that ${\bm C}={\xi^{\bm c}}$ where $\bm c$ can be obtained from (\ref{eq:genreal_boolean_c}). We follow the same notations given in (\ref{eq:ab}) in the proof of Theorem \ref{thm:GCAP_general} and consider three cases below.

{\it Case 1:} Assume $a_{\pi(1)}\neq b_{\pi(1)}$. Following a similar derivation in Case 1 in the proof of Theorem \ref{thm:GCAP_general}, we can have
\begin{equation}
\xi^{c'_{g+u_1,{i+u_2}}-c_{g,i}}+\xi^{d'_{g+u_1,{i+u_2}}-d_{g,i}}=0.
\end{equation}

{\it Case 2:} We assume $a_{\pi(1)}= b_{\pi(1)}$ and let $t$ be the smallest integer with $a_{\pi(t)}\neq b_{\pi(t)}$. The similar results can be obtained as provided in Case 2 in the proof of Theorem \ref{thm:GCAP_general}. When $1 \leq \pi(t-1)\leq n$, we have
\begin{equation}
\xi^{c'_{h,j}-c_{g,i}}+\xi^{c'_{h',j}-c_{g',i}}+\xi^{d'_{h,j}-d_{g,i}}+\xi^{d'_{h',j}-d_{g',i}}=0,
\end{equation}
and when $n< \pi(t-1)\leq n+m$, we also have
\begin{equation}
\xi^{c'_{h,j}-c_{g,i}}+\xi^{c'_{h,j'}-c_{g,i'}}+\xi^{d'_{h,j}-d_{g,i}}+\xi^{d'_{h,j'}-d_{g,i'}}=0.
\end{equation}

{\it Case 3:} Lastly, it only suffices to show that
\begin{equation}
\rho({\bm C},{\bm C'};0,0)+\rho({\bm D},{\bm D'};0,0)=0.
\end{equation}
For $1 \leq \pi(n+m)\leq n$, we have
\begin{equation}
\begin{aligned}
&\sum\limits_{g=0}^{2^{n}-1}\left(\xi^{c'_{g,i}-c_{g,i}}+\xi^{d'_{g,i}-d_{g,i}}\right)
=2\sum\limits_{g=0}^{2^{n}-1}\xi^{\frac{q}{2}g_{\pi(n+m)}}=0
\end{aligned}
\end{equation}
where $g_{\pi(n+m)}$ is the $\pi(n+m)$-th bit of the binary representation vector $(g_1,g_2,\cdots,g_n)$ of $g$. For $n < \pi(n+m)\leq m+n$, we can also obtain
\begin{equation}
\sum\limits_{i=0}^{2^{m}-1}\left(\xi^{c'_{g,i}-c_{g,i}}+\xi^{d'_{g,i}-d_{g,i}}\right)=2\sum\limits_{i=0}^{2^{m}-1}\xi^{\frac{q}{2}i_{\pi(n+m)-n}}=0
\end{equation}
where $i_{\pi(n+m)-n}$ is the $(\pi(n+m)-n)$-th bit of the binary representation vector of $i$.

Combining Case 1 to Case 3, we complete the proof.
\end{IEEEproof}
\begin{rmk}
Theorem \ref{thm:mate} includes the results in \cite[Th.7]{Pai_20} by letting $\{\pi(l),\pi(2),\cdots,\pi(n)\}=\{1,2,\cdots,n\}$ and $\{\pi(n+1),\pi(n+2),\cdots,\pi(n+m)\}=\{n+1,n+2,\cdots,n+m\}$.
Therefore, \cite[Th. 7]{Pai_20} is a special case of Theorem \ref{thm:mate}.
\end{rmk}
\begin{eg}\label{eg:mate}
Let us follow the same notations given in Example \ref{eg:GCAP_general}. Based on Theorem \ref{thm:mate}, we have an array pair $(\bm c',\bm d')=({\bm f}+{\bm z}_5,{\bm f}+{\bm z}_3+{\bm z}_5)$ given by
\[
{\bm c'}=
\begin{pmatrix}
0 & 0 & 0 & 1 & 1 & 1 & 1 & 0 \\
0 & 0 & 0 & 1 & 0 & 0 & 0 & 1 \\
0 & 0 & 1 & 0 & 1 & 1 & 0 & 1 \\
1 & 1 & 0 & 1 & 1 & 1 & 0 & 1 \\
\end{pmatrix}
~\text{and}~
{\bm d'}=
\begin{pmatrix}
0 & 1 & 0 & 0 & 1 & 0 & 1 & 1 \\
0 & 1 & 0 & 0 & 0 & 1 & 0 & 0 \\
0 & 1 & 1 & 1 & 1 & 0 & 0 & 0 \\
1 & 0 & 0 & 0 & 1 & 0 & 0 & 0 \\
\end{pmatrix}.
\]
Let $\bm C'=(\xi^{c_{g,i}'})~\text{and}~ \bm D'=(\xi^{d_{g,i}'})~\text{with} ~ 0\leq g< 4~\text{and}~ 0\leq i< 8$. Then, we list the aperiodic cross-correlations of $(\bm C,\bm C')$ and $(\bm D,\bm D')$ in (\ref{eq:R3}) and (\ref{eq:R4}). By observing the sum of the cross-correlations, $(\bm C,{\bm D})$ and $({\bm C'},{\bm D'})$ are indeed Golay complementary array mates of each other.
\begin{figure*}[ht]
\begin{equation}\label{eq:R3}
\begin{aligned}
&(\rho(\bm C, {\bm C'};u_1,u_2))_{u_{1}=-3\sim 3,u_{2}=-7\sim 7}\\
&=\left( {\begin{array}{ccccccccccccccc}
-1& 0& -1& 0& -5& 0& -1& 0& -7& 0& 1& 0& -3& 0& 1\\
-2& 0& -2& 0& -6& 0& 2& 0& 2& 0& 2&	0& 6& 0& -2\\
-3& 0& 1& 0& -3& 0& 5& 0& 7& 0&	-5&	0& -1& 0& -1\\
0& 0& 0& 0& 0& 0& 0&0&	0& 0& 0& 0&	0& 0& 0\\
1& 0& -1& 0& -7& 0& 13&0& 7& 0&	-1&	0& -1& 0& 3\\
2& 0& -6& 0& -2& 0& -2&0& -2& 0& 6&	0& 2& 0& 2\\
-1& 0& 3& 0& 3& 0& -5&0& -3& 0&	1& 0& 1& 0&	1\\
\end{array}}
\right),
\end{aligned}
\end{equation}
\end{figure*}
\begin{figure*}[ht]
\begin{equation}\label{eq:R4}
\begin{aligned}
&(\rho(\bm D, {\bm D'};u_1,u_2))_{u_{1}=-3\sim 3,u_{2}=-7\sim 7}\\
&=\left( {\begin{array}{ccccccccccccccc}
1& 0& 1& 0& 5& 0& 1& 0& 7& 0&-1& 0& 3& 0&-1\\
2& 0& 2& 0& 6& 0&-2& 0&-2& 0&-2& 0&-6& 0&2\\
3& 0&-1& 0& 3& 0&-5& 0&-7& 0& 5& 0& 1& 0&1\\
0& 0& 0& 0& 0& 0& 0& 0& 0& 0& 0& 0& 0& 0&0\\
-1& 0& 1&0& 7& 0&-13&0&-7& 0& 1& 0& 1& 0&-3\\
-2& 0& 6&0& 2& 0& 2& 0& 2& 0&-6& 0&-2& 0&-2\\
1& 0& -3&0&-3& 0& 5& 0& 3& 0&-1& 0&-1& 0&-1\\
\end{array}}
\right).
\end{aligned}
\end{equation}
\end{figure*}
\end{eg}
\subsection{GCASs Based on Generalized Boolean Functions}
In this subsection, we extend Theorem \ref{thm:GCAP_general} to propose a construction of $(N,L_1,L_2)$-GCAS with various set size $N\geq 2$.
\begin{thm}\label{thm:GCAS}
Considering nonnegative integers $n,m,k$ with $n+m\geq 2$ and $k\leq n+m$, we let nonempty sets $I_1,I_2,\cdots,I_k$ be a partition of $\{1,2,\cdots,n+m\}$. Let $t_{\alpha}$ be the order of $I_{\alpha}$ and $\pi_{\alpha}$ be a bijection from $\{1,2,\cdots,t_{\alpha}\}$ to $I_{\alpha}$ for $\alpha =1,2,\cdots,k$. If the generalized Boolean function is given by
\begin{equation}
\begin{aligned}
f= \frac{q}{2}\sum_{\alpha=1}^{k}\sum_{\beta=1}^{t_\alpha-1}z_{\pi_\alpha(\beta)}z_{\pi_\alpha(\beta+1)}+\sum_{l=1}^{n+m}p_l z_l+p_{0}
\end{aligned}
\end{equation}
where $p_l\text{'s} \in \mathbb{Z}_q$, then the array set
\begin{equation}\label{eq:set}
\begin{aligned}
&G=\left\{{\bm f}+\frac{q}{2}\sum_{\alpha=1}^{k}\lambda_{\alpha}{\bm z}_{\pi_{\alpha}(1)}:\lambda_{\alpha}\in \{0,1\}\right\}\\
\end{aligned}
\end{equation}
forms a $q$-ary $(2^k,2^n,2^m)$-GCAS.
\end{thm}
\begin{IEEEproof}
For any array $\bm c \in G$, we need to demonstrate that
\begin{equation}\label{eq:GCAS}
\begin{aligned}
\sum_{{\bf c}\in G}\sum\limits_{g=0}^{2^{n}-1}\sum\limits_{i=0}^{2^{m}-1}\left(\xi^{c_{g+u_1,{i+u_2}}-c_{g,i}}\right)&=\sum\limits_{g=0}^{2^{n}-1}\sum\limits_{i=0}^{2^{m}-1}\sum_{{\bf c}\in G}\left(\xi^{c_{g+u_1,{i+u_2}}-c_{g,i}}\right)\\
&=0
\end{aligned}
\end{equation}
for $(u_1,u_2)\neq (0,0)$. Here, we let $h=g+u_1,~j=i+u_2$ and also let $(g_1,g_2,\cdots,g_n),~(h_1,h_2,\cdots,h_n),$ $(i_1,i_2,\cdots,i_m)$, and $(j_1,j_2,\cdots,j_m)$ be the binary representation vectors of $g,~h,~i$, and $j$, respectively. By combining the binary representations of $g,~h,~i$, and $j$ as follows:
\begin{equation}
\begin{aligned}
&a_{l}=
\begin{cases}
g_{l} &\text{for}~ 1\leq l\leq n;\\
i_{l-n} &\text{for}~ n< l\leq m+n,
\end{cases}\\
&b_{l}=
\begin{cases}
h_{l} &\text{for}~ 1\leq l\leq n;\\
j_{l-n} &\text{for}~ n<l\leq m+n,
\end{cases}
\end{aligned}
\end{equation}
the proof of (\ref{eq:GCAS}) will be concise and two cases are taken into account.

{\it Case 1:} If $a_{\pi_{\alpha}(1)} \neq b_{\pi_{\alpha}(1)}$ for some $\alpha \in \{1,2,\cdots,k\}$, then for any array $\bm c \in G$, we can find an array ${\bm c}'={\bm c}+(q/2){\bm z}_{\pi_{\alpha}(1)}\in G$ satisfying
\begin{equation}
c_{h,j}-c_{g,i}-c'_{h,j}+c'_{g,i}=\frac{q}{2}(a_{\pi_{\alpha}(1)}-b_{\pi_{\alpha}(1)}) \equiv \frac{q}{2} \pmod q.
\end{equation}
Therefore, we have
\begin{equation}
\sum_{{\bf c}\in G}\xi^{c_{h,j}-c_{g,i}}=0.
\end{equation}

{\it Case 2:} Suppose that $a_{\pi_{\alpha}(1)}= b_{\pi_{\alpha}(1)}$ for all $\alpha = 1,2,\cdots,k$. We assume $a_{\pi_{\alpha}(\beta)}=b_{\pi_{\alpha}(\beta)}$ for $\alpha = 1,\cdots,\hat{\alpha}-1$ and $\beta=1,2,\cdots, m_{\alpha}$. Besides, we assume that $\hat{\beta}$ is the smallest number satisfying $a_{\pi_{\hat{\alpha}}(\hat{\beta}-1)}\neq b_{\pi_{\hat{\alpha}}(\hat{\beta}-1)}$. Let $a'_{\pi_{\hat{\alpha}}(\hat{\beta}-1)}=1-a_{\pi_{\hat{\alpha}}(\hat{\beta}-1)}$ and $b'_{\pi_{\hat{\alpha}}(\hat{\beta}-1)}=1-b_{\pi_{\hat{\alpha}}(\hat{\beta}-1)}$. Following the similar argument as mentioned in Case 2 in the proof of Theorem \ref{thm:GCAP_general}, we have
\begin{equation}
\sum_{{\bf c}\in G} \xi^{c_{h,j}-c_{g,i}}+\xi^{c_{h',j}-c_{g',i}}=0,
\end{equation}
for $1 \leq \pi_{\hat{\alpha}}(\hat{\beta}-1)\leq n$ and
\begin{equation}
\sum_{{\bf c}\in G} \xi^{c_{h,j}-c_{g,i}}+\xi^{c_{h,j'}-c_{g,i'}}=0.
\end{equation}
for $n< \pi_{\hat{\alpha}}(\hat{\beta}-1)\leq n+m$.

According to the above two cases, the equality in (\ref{eq:GCAS}) holds.
\end{IEEEproof}
\begin{eg}\label{eg:GCAS}
Let us take $q=2,~ n=2,$ and $m=3$ for example. According to Theorem \ref{thm:GCAS}, we let $I_1=\{4,2,5\}, ~I_2=\{1,3\},~\pi_{1}(1)=4,~\pi_{1}(2)=2,~\pi_{1}(3)=5,\pi_{2}(1)=1,~\pi_{2}(2)=3$, and the Boolean function
$f=z_4z_2+z_2z_5+z_1z_3$, which can be represented by $f=x_2y_2+y_2x_3+y_1x_1$ according to (\ref{eq:z}). The set $G=\{\bm f,\bm f+{\bm z}_4,\bm f+{\bm z}_1,{\bm f}+{\bm z}_4+{\bm z}_1\}$ forms a $q$-ary $(4,4,8)$-GCAS where
\[
{\bm f}=
\begin{pmatrix}
0 & 0 & 0 & 0 & 0 & 0 & 0 & 0 \\
0 & 1 & 0 & 1 & 0 & 1 & 0 & 1 \\
0 & 0 & 1 & 1 & 1 & 1 & 0 & 0 \\
0 & 1 & 1 & 0 & 1 & 0 & 0 & 1 \\
\end{pmatrix},
~{\bm f+{\bm z}_4}=
\begin{pmatrix}
0 & 0 & 1 & 1 & 0 & 0 & 1 & 1 \\
0 & 1 & 1 & 0 & 0 & 1 & 1 & 0 \\
0 & 0 & 0 & 0 & 1 & 1 & 1 & 1 \\
0 & 1 & 0 & 1 & 1 & 0 & 1 & 0 \\
\end{pmatrix},
\]
\[
{\bm f+{\bm z}_1}=
\begin{pmatrix}
0 & 0 & 0 & 0 & 0 & 0 & 0 & 0 \\
1 & 0 & 1 & 0 & 1 & 0 & 1 & 0 \\
0 & 0 & 1 & 1 & 1 & 1 & 0 & 0 \\
1 & 0 & 0 & 1 & 0 & 1 & 1 & 0 \\
\end{pmatrix},
~{\bm f+{\bm z}_4+{\bm z}_1}=
\begin{pmatrix}
0 & 0 & 1 & 1 & 0 & 0 & 1 & 1 \\
1 & 0 & 0 & 1 & 1 & 0 & 0 & 1 \\
0 & 0 & 0 & 0 & 1 & 1 & 1 & 1 \\
1 & 0 & 1 & 0 & 0 & 1 & 0 & 1 \\
\end{pmatrix}.
\]
According to the mapping in (\ref{eq:z}), we have $z_1=x_1,~z_2=x_2,~z_3=y_1,~z_4=y_2$, and $z_5=y_3$. Therefore, we have ${\bm F}_1=\xi^{\bms f}=(\xi^{f_{g,i}}),~{\bm F}_2=\xi^{\bms f+{\bms z_4}}=(\xi^{f_{g,i}+i_2}),~{\bm F}_3=\xi^{\bms f+{\bms z_1}}=(\xi^{f_{g,i}+g_1}),~\text{and}~{\bm F}_4=\xi^{\bms f+{\bms z_4}+{\bms z_1}}=(\xi^{f_{g,i}+i_2+g_1})$, where $0\leq g< 4,~ 0\leq i< 8$ and $g=\sum_{s=1}^{2}g_{s}2^{s-1},~i=\sum_{l=1}^{3}i_{l}2^{l-1}$. Their aperiodic autocorrelations are given, respectively, in (\ref{eq:R5}) to (\ref{eq:R8}). We can observe that the sums of their autocorrelations are all zero for $(u_1,u_2)\neq (0,0)$.
\begin{figure*}[ht]
\begin{equation}\label{eq:R5}
\begin{aligned}
&(\rho(\bm F_1;u_1,u_2))_{u_{1}=-3\sim 3,u_{2}=-7\sim 7}\\
&=\left( {\begin{array}{ccccccccccccccc}
-1&	0&	1&	0&	1&	0&	-1&	0&	1&	0&	-1&	0&	-1&	0&	1\\
0&	4&	0&	0&	0&	-4&	0&	0&	0&	-4&	0&	0&	0&	4&	0\\
-1&	0&	1&	0&	-3&	0&	-5&	0&	5&	0&	3&	0&	-1&	0&	1\\
0&	8&	0&	0&	0&	8&	0&	32&	0&	8&	0&	0&	0&	8&	0\\
1&	0&	-1&	0&	3&	0&	5&	0&	-5&	0&	-3&	0&	1&	0&	-1\\
0&	4&	0&	0&	0&	-4&	0&	0&	0&	-4&	0&	0&	0&	4&	0\\
1&	0&	-1&	0&	-1&	0&	1&	0&	-1&	0&	1&	0&	1&	0&	-1
\end{array}}
\right),
\end{aligned}
\end{equation}
\end{figure*}
\begin{figure*}[ht]
\begin{equation}\label{eq:R6}
\begin{aligned}
&(\rho(\bm F_2;u_1,u_2))_{u_{1}=-3\sim 3,u_{2}=-7\sim 7}\\
&=\left( {\begin{array}{ccccccccccccccc}
1&	0&	-1&	0&	-1&	0&	1&	0&	-1&	0&	1&	0&	1&	0&	-1\\
0&	-4&	0&	0&	0&	4&	0&	0&	0&	4&	0&	0&	0&	-4&	0\\
1&	0&	-1&	0&	3&	0&	-11&0&	11&	0&	-3&	0&	1&	0&	-1\\
0&	-8&	0&	0&	0&	-8&	0&	32&	0&	-8&	0&	0&	0&	-8&	0\\
-1&	0&	1&	0&	-3&	0&	11&	0&	-11&0&	3&	0&	-1&	0&	1\\
0&	-4&	0&	0&	0&	4&	0&	0&	0&	4&	0&	0&	0&	-4&	0\\
-1&	0&	1&	0&	1&	0&	-1&	0&	1&	0&	-1&	0&	-1&	0&	1
\end{array}}
\right),
\end{aligned}
\end{equation}
\end{figure*}

\begin{equation}\label{eq:R7}
\begin{aligned}
&(\rho(\bm F_3;u_1,u_2))_{u_{1}=-3\sim 3,u_{2}=-7\sim 7}\\
&=\left( {\begin{array}{ccccccccccccccc}
1&	0&	-1&	0&	-1&	0&	1&	0&	-1&	0&	1&	0&	1&	0&	-1\\
0&	4&	0&	0&	0&	-4&	0&	0&	0&	-4&	0&	0&	0&	4&	0\\
1&	0&	-1&	0&	3&	0&	5&	0&	-5&	0&	-3&	0&	1&	0&	-1\\
0&	8&	0&	0&	0&	8&	0&	32&	0&	8&	0&	0&	0&	8&	0\\
-1&	0&	1&	0&	-3&	0&	-5&	0&	5&	0&	3&	0&	-1&	0&	1\\
0&	4&	0&	0&	0&	-4&	0&	0&	0&	-4&	0&	0&	0&	4&	0\\
-1&	0&	1&	0&	1&	0&	-1&	0&	1&	0&	-1&	0&	-1&	0&	1
\end{array}}
\right),
\end{aligned}
\end{equation}
\begin{equation}\label{eq:R8}
\begin{aligned}
&(\rho(\bm F_4;u_1,u_2))_{u_{1}=-3\sim 3,u_{2}=-7\sim 7}\\
&=\left( {\begin{array}{ccccccccccccccc}
-1&	0&	1&	0&	1&	0&	-1&	0&	1&	0&	-1&	0&	-1&	0&	1\\
0&	-4&	0&	0&	0&	4&	0&	0&	0&	4&	0&	0&	0&	-4&	0\\
-1&	0&	1&	0&	-3&	0&	11&	0&	-11&0&	3&	0&	-1&	0&	1\\
0&	-8&	0&	0&	0&	-8&	0&	32&	0&	-8&	0&	0&	0&	-8&	0\\
1&	0&	-1&	0&	3&	0&	-11&0&	11&	0&	-3&	0&	1&	0&	-1\\
0&	-4&	0&	0&	0&	4&	0&	0&	0&	4&	0&	0&	0&	-4&	0\\
1&	0&	-1&	0&	-1&	0&	1&	0&	-1&	0&	1&	0&	1&	0&	-1
\end{array}}
\right).
\end{aligned}
\end{equation}
\end{eg}

Compared to \cite{GCAP_Zeng_10}, Theorem \ref{thm:GCAS} constructs aperiodic GCASs and does not require any perfect array as a kernel. 2-D GCASs can be generated directly from generalized Boolean functions.
\section{Conclusion}\label{sec:conclusion}
In this paper, novel constructions of 2-D GCAPs, 2-D GCASs, and 2-D Golay complementary array mates based on 2-D generalized Boolean functions have been proposed. First, we give the basic construction of 2-D GCAPs of size $2^n \times 2^{m}$ from Boolean functions in Theorem \ref{thm:GCAP}. Then, we propose extended constructions of 2-D GCAPs and 2-D Golay complementary array mates, respectively, in Theorem \ref{thm:GCAP_general} and Theorem \ref{thm:mate}, which can include the results in \cite[Th.6]{Pai_20} and \cite[Th.7]{Pai_20} as special cases. By adopting the proper mapping given in (\ref{eq:z}), Theorem \ref{thm:GCAP_general} presents an elegant and explicit expression for 2-D GCAPs in terms of 2-D Boolean functions. In addition, the upper bounds on row sequence PAPR and column sequence PAPR of the proposed 2-D GCAPs are derived in Corollary \ref{crly:row_general} and Corollary \ref{crly:col_general}. Moreover, we further propose $(2^k,2^n,2^m)$-GCASs based on generalized Boolean functions from Theorem \ref{thm:GCAS}. This is, to the authors' knowledge, the first construction method of 2-D GCASs based on Boolean functions. The proposed constructions are all direct constructions without using other existing 1-D sequences or 2-D arrays as kernels.

Although Theorem \ref{thm:GCAS} can provide a direct construction of 2-D GCASs, the array sizes of each constituent array are limited to $2^n \times 2^m$. Therefore, possible future research is to investigate the construction of 2-D GCASs with various array sizes. Besides, an extension to 2-D CCCs and 2-D Z-Complementary array pairs/sets \cite{Pai_19,Das_20,Pai_21} based on generalized Boolean functions is also an interesting topic.

%
%

\bibliographystyle{IEEEtran}
\bibliography{IEEEabrv,2D_ZCP}

\begin{thebibliography}{10}
\providecommand{\url}[1]{#1}
\csname url@samestyle\endcsname
\providecommand{\newblock}{\relax}
\providecommand{\bibinfo}[2]{#2}
\providecommand{\BIBentrySTDinterwordspacing}{\spaceskip=0pt\relax}
\providecommand{\BIBentryALTinterwordstretchfactor}{4}
\providecommand{\BIBentryALTinterwordspacing}{\spaceskip=\fontdimen2\font plus
\BIBentryALTinterwordstretchfactor\fontdimen3\font minus
  \fontdimen4\font\relax}
\providecommand{\BIBforeignlanguage}[2]{{%
\expandafter\ifx\csname l@#1\endcsname\relax
\typeout{** WARNING: IEEEtran.bst: No hyphenation pattern has been}%
\typeout{** loaded for the language `#1'. Using the pattern for}%
\typeout{** the default language instead.}%
\else
\language=\csname l@#1\endcsname
\fi
#2}}
\providecommand{\BIBdecl}{\relax}
\BIBdecl

\bibitem{Golay}
M.~J.~E. Golay, ``Complementary series,'' \emph{{IRE} Trans. Inf. Theory}, vol.
  IT-7, pp. 82--87, Apr. 1961.

\bibitem{Tseng72}
C.-C. Tseng and C.~L. Liu, ``Complementary sets of sequences,'' \emph{{IEEE}
  Trans. Inf. Theory}, vol. IT-18, pp. 644--652, Sep. 1972.

\bibitem{CS_CE}
P.~Spasojevic and C.~N. Georghiades, ``Complementary sequences for {ISI}
  channel estimation,'' \emph{{IEEE} Trans. Inf. Theory}, vol.~47, no.~3, pp.
  1145--1152, Mar. 2001.

\bibitem{CS_sync}
J.~M. Groenewald and B.~T. Maharaj, ``{MIMO} channel synchronization using
  {Golay} complementary pairs,'' in \emph{Proc.\ AFRICON 2007}, Windhoek, South
  Africa, Sep. 2007, pp. 1--5.

\bibitem{Chen_CDMA}
H.-H. Chen, J.-F. Yeh, and N.~Suehiro, ``A multicarrier {CDMA} architecture
  based on orthogonal complete complementary codes for new generations of
  wideband wireless communications,'' \emph{{IEEE} Commun. Mag.}, vol.~39, pp.
  126--134, Oct. 2001.

\bibitem{Golay_power1}
S.~Boyd, ``Multitone signals with low crest factor,'' \emph{{IEEE} Trans.
  Circuits Syst.}, vol. CAS-33, no.~10, pp. 1018--1022, Oct. 1986.

\bibitem{Nee2}
R.~van Nee, ``{OFDM} codes for peak-to-average power reduction and error
  correction,'' in \emph{Proc. {IEEE} Global Telecommun. Conf.}, London, U.K.,
  Nov. 1996, pp. 740--744.

\bibitem{Golay_RM}
J.~A. Davis and J.~Jedwab, ``Peak-to-mean power control in {OFDM}, {Golay}
  complementary sequences, and {Reed-Muller} codes,'' \emph{{IEEE} Trans. Inf.
  Theory}, vol.~45, no.~7, pp. 2397--2417, Nov. 1999.

\bibitem{Paterson_00}
K.~G. Paterson, ``Generalized {Reed-Muller} codes and power control in {OFDM}
  modulation,'' \emph{{IEEE} Trans. Inf. Theory}, vol.~46, no.~1, pp. 104--120,
  Jan. 2000.

\bibitem{Dymond_92}
M.~Dymond, ``Barker arrays: existence, generalization and alternatives,''
  \emph{Ph.D. thesis, University of London}, 1992.

\bibitem{GCAP_Mtsufuji_04}
S.~Matsufuji, R.~Shigemitsu, Y.~Tanada, and N.~Kuroyanagi, ``Construction of
  complementary arrays,'' in \emph{Proc.\ Joint IST Workshop on Mobile Future
  and Symp. on Trends in Commun.}, Bratislava, Slovakia, Oct. 2004, pp. 78--81.

\bibitem{GCAP_Jedwab_07}
J.~Jedwab and M.~G. Parker, ``Golay complementary array pairs,'' \emph{Designs,
  Codes and Cryptography}, vol.~44, no. 1-3, p. 209–216, Sep. 2007.

\bibitem{GCAP_Fiedler_08}
F.~Fiedler, J.~Jedwab, and M.~G. Parker, ``A multi-dimensional approach for the
  construction and enumeration of {Golay} complementary sequences,'' \emph{J.
  Combinatorial Theory (Series A)}, vol. 115, no.~5, pp. 753--776, Jul. 2008.

\bibitem{GCAP_Zeng_10}
F.~Zeng and Z.~Zhang, ``Two dimensional periodic complementary array sets,'' in
  \emph{Proc.\ IEEE Int. Conf. on Wireless Commun., Netw. and Mobile Comput.},
  Chengdu, China, Sep. 2010, pp. 1--4.

\bibitem{Jiang_19}
Y.~Jiang, F.~Li, X.~Wang, and J.~Li, ``Autocorrelation complementary
  matrices,'' in \emph{Proc. 53rd Asilomar Conf. on Signals, Systems, and
  Computers}, Pacific Grove, CA, Nov. 2019, pp. 1--5.

\bibitem{Weathers_1983}
G.~Weathers and E.~M. Holliday, ``Group-complementary array coding for radar
  clutter rejection,'' \emph{IEEE Trans. Aerospace and Electronic Systems},
  vol. AES-19, no.~3, pp. 369--379, May 1983.

\bibitem{Golomb_1982}
S.~W. Golomb and H.~Taylor, ``Two-dimensional synchronization patterns for
  minimum ambiguity,'' \emph{{IEEE} Trans. Inf. Theory}, vol.~28, no.~4, pp.
  600--604, Jul. 1982.

\bibitem{Hershey_1983}
J.~E. Hershey and R.~Yarlagadda, ``Two-dimensional synchronisation,''
  \emph{Electronics Letters}, vol.~19, no.~19, pp. 801 -- 803, Sep. 1983.

\bibitem{2DCDMA_2004}
M.~Turcsány and P.~Farkaš, ``New {2D-MC-DS-SS-CDMA} techniques based on
  two-dimensional orthogonal complete complementary codes,'' in \emph{Proc.
  Multi-Carrier Spread-Spectrum}, Dordrecht, Netherlands, Jan. 2004, pp.
  49--56.

\bibitem{2D_CCC}
P.~Farkaš and M.~Turcsány, ``Two-dimensional orthogonal complete
  complementary codes,'' in \emph{Proc.\ Joint IST Workshop on Mobile Future
  and Symp. on Trends in Commun.}, Bratislava, Slovakia, Oct. 2003, pp. 1--5.

\bibitem{AAECC2006}
{C.-Y.\ Chen, C.-H.\ Wang, and C.-C.\ Chao}, ``Complementary sets and
  {Reed-Muller} codes for peak-to-average power ratio reduction in {OFDM},'' in
  \emph{Proc.\ 16th\ Int.\ Symp.\ AAECC, LNCS 3857}, Las Vegas, NV, Feb. 2006,
  pp. 317--327.

\bibitem{Super_16}
C.-Y. Chen, ``Complementary sets of non-power-of-two length for peak-to-average
  power ratio reduction in {OFDM},'' \emph{{IEEE} Trans. Inf. Theory}, vol.~62,
  no.~12, pp. 7538--7545, Dec. 2016.

\bibitem{Super_172}
------, ``A new construction of {Golay} complementary sets of non-power-of-two
  length based on {B}oolean functions,'' in \emph{Proc.\ IEEE Wireless Commun.
  and Netw. Conf.}, San Francisco, CA, Mar. 2017, pp. 1--6.

\bibitem{Super_18}
------, ``A novel construction of complementary sets with flexible lengths
  based on {Boolean} functions,'' \emph{{IEEE} Commun. Lett.}, vol.~22, no.~2,
  pp. 260--263, Feb. 2018.

\bibitem{schmidt}
K.-U. Schmidt, ``Complementary sets, generalized {Reed-Muller} codes, and power
  control for {OFDM},'' \emph{{IEEE} Trans. Inf. Theory}, vol.~53, no.~2, pp.
  808--814, Feb. 2007.

\bibitem{MutualGCS_2008}
A.~Rathinakumar and A.~K. Chaturvedi, ``Complete mutually orthogonal {Golay}
  complementary sets from {Reed-Muller} codes,'' \emph{{IEEE} Trans. Inf.
  Theory}, vol.~54, pp. 1339--1346, Mar. 2008.

\bibitem{Chen08}
C.-Y. Chen, C.-H. Wang, and C.-C. Chao, ``Complete complementary codes and
  generalized {Reed-Muller} codes,'' \emph{{IEEE} Commun. Lett.}, vol.~12, pp.
  849--851, Nov. 2008.

\bibitem{Liu_2014}
Z.~Liu, Y.~L. Guan, and U.~Parampalli, ``New complete complementary codes for
  peak-to-mean power control in multi-carrier {CDMA},'' \emph{{IEEE} Trans.
  Commun.}, vol.~62, pp. 1105--1113, Mar. 2014.

\bibitem{Wu_20}
S.-W. Wu, C.-Y. Chen, and Z.~Liu, ``How to construct mutually orthogonal
  complementary sets with non-power-of-two lengths?'' \emph{{IEEE} Trans. Inf.
  Theory}, 2020, {E}arly Access.

\bibitem{Pai_202}
C.-Y. Pai, S.-W. Wu, and C.-Y. Chen, ``{Z}-complementary pairs with flexible
  lengths from generalized {B}oolean functions,'' \emph{{IEEE} Commun. Lett.},
  vol.~24, no.~6, pp. 1183 -- 1187, Jun. 2020.

\bibitem{odd-ZCP}
Z.~Liu, U.~Parampalli, and Y.~L. Guan, ``Optimal odd-length binary
  {Z}-complementary pairs,'' \emph{{IEEE} Trans. Inf. Theory}, vol.~60, no.~9,
  pp. 5768--5781, Sep. 2014.

\bibitem{even-ZCP}
------, ``On even-period binary {Z}-complementary pairs with large {ZCZs},''
  \emph{{IEEE} Signal Process. Lett.}, vol.~21, no.~3, pp. 284--287, Mar. 2014.

\bibitem{Xie_18}
C.~Xie and Y.~Sun, ``Constructions of even-period binary {Z}-complementary
  pairs with large {ZCZs},'' \emph{IEEE Signal Process. Lett.}, vol.~25, no.~8,
  pp. 1141--1145, Aug. 2018.

\bibitem{Adhikary_20}
A.~R. Adhikary, P.~Sarkar, and S.~Majhi, ``A direct construction of $q$-ary
  even length {Z}-complementary pairs using generalized {Boolean} functions,''
  \emph{{IEEE} Signal Process. Lett.}, vol.~27, pp. 146 -- 150, 2020.

\bibitem{Shen_19}
B.~Shen, Y.~Yang, Z.~Zhou, P.~Fan, and Y.~L. Guan, ``New optimal binary
  {Z}-complementary pairs of odd length $2^m + 3$,'' \emph{{IEEE} Signal
  Process. Lett.}, vol.~26, no.~12, pp. 1931--1934, Dec. 2019.

\bibitem{Wu_18}
S.-W. Wu and C.-Y. Chen, ``Optimal {Z}-complementary sequence sets with good
  peak-to-average power-ratio property,'' \emph{IEEE Signal Process. Lett.},
  vol.~25, no.~10, pp. 1500--1504, Oct. 2018.

\bibitem{Sarkar_19}
P.~Sarkar, S.~Majhi, and Z.~Liu, ``Optimal {Z}-complementary code set from
  generalized {Reed-Muller} codes,'' \emph{{IEEE} Trans. Commun.}, vol.~67,
  no.~3, pp. 1783--1796, Mar. 2019.

\bibitem{Luke_85}
H.~D. L{\"u}ke, ``Sets of one and higher dimensional {W}elti codes and
  complementary codes,'' \emph{IEEE Trans. Aerosp. Electron. Syst.}, vol.
  AES-21, no.~2, pp. 170--179, Apr. 1985.

\bibitem{2D_CCC_CDMA}
P.~Farkaš, M.~Turcsány, and H.~Bali, ``Application of 2{D} complete
  complementary orthogonal codes in {2D-MC-SS-CDMA},'' in \emph{Proc.\ Int.\
  Symp. Wireless Personal Multimedia Commun.}, Abano Terme, Italy, Sep. 2004,
  pp. 1--5.

\bibitem{Pai_20}
C.-Y. Pai and C.-Y. Chen, ``Constructions of two-dimensional {Golay}
  complementary array pairs based on generalized {Boolean} functions,'' in
  \emph{Proc.\ {IEEE} Int.\ Symp. Inf. Theory}, Los Angeles, CA, Jun. 2020, pp.
  2931--2935.

\bibitem{Li_15}
Y.~Li and C.~Xu, ``{ZCZ} aperiodic complementary sequence sets with low column
  sequence {PMEPR},'' \emph{{IEEE} Commun. Lett.}, vol.~19, no.~8, pp.
  1303--1306, Aug. 2015.

\bibitem{Pai_19}
C.-Y. Pai, Y.-T. Ni, Y.-C. Liu, M.-H. Kuo, and C.-Y. Chen, ``Constructions of
  two-dimensional binary {Z}-complementary array pairs,'' in \emph{Proc.\
  {IEEE} Int.\ Symp. Inf. Theory}, Paris, France, Jul. 2019, pp. 2264--2268.

\bibitem{Das_20}
S.~Das and S.~Majhi, ``Two-dimensional {Z}-complementary array code sets based
  on matrices of generating polynomials,'' \emph{{IEEE} Trans. Signal
  Process.}, {Early} Access.

\bibitem{Pai_21}
C.-Y. Pai, Y.-T. Ni, and C.-Y. Chen, ``Two-dimensional binary {Z}-complementary
  array pairs,'' \emph{{IEEE} Trans. Inf. Theory}, accepted for publication.

\end{thebibliography}
\IEEEtriggeratref{3}

\end{document}